\documentclass[12pt]{article}
\usepackage{epsfig}
\begin{document}
\renewcommand{\theequation}{\thesection.\arabic{equation}}
\title{From the  WZWN Model to the Liouville Equation: Exact String
Dynamics in Conformally
Invariant AdS Background}
\author{ A.L. Larsen\thanks{Department of
Physics, University of Odense,
Campusvej 55, 5230 Odense M, Denmark.}
and
N. S\'{a}nchez\thanks{Observatoire de Paris,
DEMIRM. Laboratoire Associ\'{e} au CNRS
UA 336, Observatoire de Paris et
\'{E}cole Normale Sup\'{e}rieure. 61, Avenue
de l'Observatoire, 75014 Paris, France.}}
\maketitle
\begin{abstract}
It has been known for some time that the SL(2,R) WZWN model reduces
to
Liouville theory.
Here we give a direct and physical derivation of this result based on
the
classical
string equations of motion and the proper string size. This allows us
to
extract precisely the physical
effects of the metric and antisymmetric tensor, respectively, on the
{\it exact} string dynamics in the SL(2,R)
background. The general solution to the proper string size is also
found.
We show that the
antisymmetric tensor (corresponding to conformal invariance)
generally gives
rise to repulsion, and it precisely cancels the dominant attractive
term arising from the metric.

Both the sinh-Gordon and the cosh-Gordon sectors of the string
dynamics in
non-conformally
invariant AdS spacetime reduce here to the Liouville equation (with
different signs of the
potential), while the original Liouville sector reduces to the free
wave
equation. Only the
very large classical string size is affected by the torsion. Medium
and
small size string
behaviours are unchanged.

We also find illustrative classes of string solutions in the SL(2,R)
background: dynamical
closed as well as stationary open spiralling strings, for which the
effect
of torsion is somewhat
like the effect of rotation in the metric. Similarly, the string
solutions in
the 2+1 BH-AdS background  with torsion and angular momentum are
fully
analyzed.
\end{abstract}
\newpage
\section{Introduction}
\setcounter{equation}{0}
In this paper, we consider the exact classical string dynamics in the
conformally
invariant background corresponding to the SL(2,R) WZWN model. This
background is locally $2+1$-dimensional Anti de Sitter
spacetime with non-vanishing parallelizing torsion.

Many mathematical aspects of the SL(2,R) WZWN model have been
discussed in
the literature (see for
instance Refs.\cite{bal,pet,nem,hwa}). In particular, it has been
known for
some time that
the SL(2,R) WZWN
model  reduces
to Liouville theory  (for a review of the different methods, see
\cite{rai}
and references given therein).
However, the physical aspects have still not really been
extracted so far. The purpose of
this paper is to investigate directly the physical effects of the
conformal
invariance on the generic exact classical string dynamics.

The conformal invariance of the SL(2,R) WZWN model
is expressed as the  torsion becoming parallelizing. Thus we consider
the
string
equations of motion in a
background consisting of the  Anti de Sitter (AdS) metric plus an
antisymmetric tensor representing
the parallelizing torsion. Using the reduction method of
\cite{san,all}, we
obtain
directly from the classical
equations of motion a simple differential equation, the Liouville
equation,
for the
fundamental quadratic form
$\alpha(\tau,\sigma)$, which determines the proper string size.   By
comparing with  analogues results
obtained in AdS but without torsion \cite{san,all,mik,nes},
we can then precisely extract the physical
effects of the conformal invariance
on the {\it exact} dynamics of classical strings. We also compare
with the
results of
\cite{vega} where, among other
things, the effect of the conformal invariance was analysed, but for
particular string
configurations and for perturbative
string solutions.

One essential point in this paper is the parametrization of the
string
equations of motion
and constraints in terms of the proper string size. Then, associated
potentials $V(\alpha)$
can be defined, and {\it generic} properties of the exact string
dynamics
can be extracted
directly from the reduced equations of motion and potentials (without
need
of any solution).
Previously \cite{all}, we have shown that the exact string dynamics
in the
non-conformally
invariant AdS spacetime (without torsion), reduces to three different
equations:
sinh-Gordon, cosh-Gordon and Liouville equation, and all three must
be
considered in order
to cover the generic string evolution.

In this paper we show that this reduction procedure beautifully
generalizes
and {\it
simplifies} in the presence of torsion, corresponding to conformal
invariance. In the
conformally invariant AdS background, the presence of torsion leads
to a
{\it precise}
cancellation of the term $+\mbox{exp}(\alpha)$ in the potentials
$\cosh(\alpha)$,
$\sinh(\alpha)$ and
$\mbox{exp}(\alpha)$ of the reduced equations. Thus, when including
the
torsion, the
original sinh-Gordon and cosh-Gordon sectors reduce to the Liouville
equation (with
different signs of the potential), while the original Liouville
sector
reduces to the free
wave equation (see Figs. 1, 2). Torsion generally produces a
repulsive term
$-\mbox{exp}(\alpha)$, which precisely cancels the dominant
attractive term
arising from
gravity. As a consequence, only the very large classical string size
behaviour is
affected by the torsion. Most of the string behaviour (medium and
small
string size
behaviour) is unchanged.

We also find in this paper  illustrative classes of string solutions
in
the conformally
invariant AdS background. The ansatz we have introduced in
Ref.\cite{all}
is applied here to
this case. Dynamical closed strings as well as stationary open
infinitely
long strings are
described. Here in the presence of torsion, the mathematics
simplifies
considerably; the
solutions are expressed in closed form in terms of trigonometric and
hyperbolic functions
(in the non-conformally invariant case, the solutions generally
involved
elliptic functions
\cite{all}). Similarly, we find the string solutions in the 2+1
dimensional
black hole anti de Sitter spacetime (BH-AdS) with torsion, and
compare with the
case of vanishing torsion.

It must be noticed that, in the absence of torsion, stationary
strings in
AdS spacetime are
of "hanging string" type (that is, their shapes are simple
generalizations
of the shape of a
rope hanging in a constant Newtonian potential). In the presence of
torsion, these
configurations become of "spiralling string" type, and
asymptotically, they
are standard
logaritmic spirals (see Figs. 3, 4). The effect of torsion on
stationary
strings in the AdS
background thus appears quite similar to the effect of rotation in
the
Kerr-Newman spacetime
\cite{zel}.

This paper is organized as follows: In Section 2 we perform the
reduction
of the WZWN model
to the Liouville equation, in terms of the string equations of motion
and
constraints and the
proper string size, and we analyse the generic features of the exact
string
dynamics in the
AdS background with torsion. In Section 3 we deal with particularly
illustrative examples of
string configurations and precise effects of the torsion and
conformal
invariance in this
background, using a parametrization corresponding to {\it global} AdS
spacetime. In Section 4, we discuss the analogous results obtained
using a parametrization corresponding to the 2+1 dimensional BH-AdS
spacetime
with torsion. Conclusions and remarks are given in Section 5.
\section{Reduction of the WZWN Model to the Liouville Equation}
\setcounter{equation}{0}
Our starting point is the sigma-model action including the WZWN term
at
level
$k$ \cite{wit}:
\begin{equation}
S_{\sigma}=-\frac{k}{4\pi}\int_{M} d\tau
d\sigma\;\eta^{\alpha\beta}\mbox{Tr}[
g^{-1}\partial_\alpha g\;g^{-1}\partial_\beta g]-
\frac{k}{6\pi}\int_{B} \mbox{Tr}[
g^{-1}dg\wedge g^{-1}dg\wedge g^{-1}dg].
\end{equation}
Here $M$ is the boundary of the manifold $B$, and $g$ is a
group-element of
$SL(2,R)$:
\begin{equation}
g=\left( \begin{array}{cc} a & c \\
-d & b \end{array}\right) ;\;\;\;\;\;\;\;\;\ ab+cd=1.
\end{equation}
Then, the action (2.1) takes the form \cite{hor}:
\begin{equation}
S_{\sigma}=-\frac{k}{2\pi}\int_{M} d\tau
d\sigma\;[\dot{a}\dot{b}-a'b'+
\dot{c}\dot{d}-c'd']-\frac{k}{\pi}\int_{M} d\tau d\sigma\;
\log (c) [\dot{a}b'-a'\dot{b}],
\end{equation}
where dot and prime denote derivative with respect to $\tau$ and
$\sigma$,
respectively. Let us introduce new coordinates
$(X,Y,U,T)$:
\begin{equation}
a=H(U+X),\;\;\;\;\;\;b=H(U-X),\;\;\;\;\;\;c=H(T-Y),\;\;\;\;\;\;d=H(T+Y
),
\end{equation}
where $H$ is a constant (the Hubble constant). Then we get from
(2.2):
\begin{equation}
X^2+Y^2-U^2-T^2=-\frac{1}{H^2},
\end{equation}
which is the standard embedding equation for the 2+1 AdS spacetime.
Using a Lagrange
multiplier $\lambda$ to
incorporate
the condition (2.5), the action becomes:
\begin{eqnarray}
S_{\sigma}=-\frac{k H^2}{2\pi}\int_{M} d\tau
d\sigma\;[\dot{U}^2-U'^2+\dot{T}^2
-T'^2-\dot{X}^2 +X'^2-\dot{Y}^2 +Y'^2
+4\lambda(-T^2\nonumber\\
-U^2+X^2+Y^2+H^{-2})+4(\dot{X}U'-X'\dot{U})\log
(H(T-Y))].
\nonumber\\
\end{eqnarray}
It is also convenient to introduce the dimensionless 4-vector $q^\mu$
and
metric $\eta_{\mu\nu}$ in the
4-dimensional embedding spacetime:
\begin{equation}
q^\mu=H(T,U,X,Y),\;\;\;\;\;\;\;\;\eta_{\mu\nu}=\mbox{diag}
(-1,-1,1,1),
\end{equation}
as well as world-sheet light-cone coordinates:
\begin{equation}
\sigma^\pm=\tau\pm\sigma.
\end{equation}
It is then straightforward to show that the classical equations of
motion
corresponding to the action
(2.6) reduce to:
\begin{equation}
q^\mu_{+-}+e^\alpha q^\mu+\epsilon^\mu_{\;\;\nu\rho\sigma}
q^\nu q^\rho_+
q^\sigma_-=0,\;\;\;\;\;\;\;\;(\mu=0,1,2,3),
\end{equation}
where we introduced the fundamental quadratic form
$\alpha(\tau,\sigma)$:
\begin{equation}
e^\alpha=-\eta_{\mu\nu}q^\mu_+ q^\nu_-,
\end{equation}
and the antisymmetric tensor $\epsilon^\mu_{\;\;\nu\rho\sigma}$
corresponding to the metric (2.7):
\begin{equation}
\epsilon^{0123}=1,\;\;\;\;(\mbox{antisymmetric}).
\end{equation}
The equations of motion (2.9) should, as usual, be supplemented by
the
string constraints:
\begin{equation}
\eta_{\mu\nu}q^\mu_\pm q^\nu_{\pm}=0,
\end{equation}
and the embedding normalization condition (2.5):
\begin{equation}
\eta_{\mu\nu}q^\mu q^\nu=-1.
\end{equation}
Notice that $\alpha$ determines the proper string size, as follows
from the
induced metric on the
world-sheet:
\begin{equation}
dS^2=\frac{1}{H^2}\eta_{\mu\nu}dq^\mu dq^\nu
=\frac{2}{H^2}e^\alpha\;(-d\tau^2+d\sigma^2).
\end{equation}
It is convenient to introduce the basis:
\begin{equation}
{\cal U}=
\{ q^\mu, q^\mu_+,q^\mu_-,l^\mu\};
\;\;\;\;\;\;\;\;\;\;\;\;\;
l^\mu\equiv
e^{-\alpha} \epsilon^\mu_{\;\;\rho\sigma\delta}
q^\rho q^\sigma_+ q^\delta_-,
\end{equation}
\begin{equation}
\eta_{\mu\nu}l^\mu l^\nu=1.
\end{equation}
The second derivatives of $q^\mu,$
expressed in the basis
${\cal U},$ are
given by:
\begin{equation}
q^\mu_{++}=\alpha_+ q^\mu_+ +u
l^\mu,\;\;\;\;
q^\mu_{--}=\alpha_- q^\mu_- +v
l^\mu,\;\;\;\;
q^\mu_{+-}=-e^{\alpha}(q^\mu+l^\mu),
\end{equation}
where the functions $u$ and $v$ are
implicitly
defined by:
\begin{equation}
u\equiv
\eta_{\mu\nu}q^{\mu}_{++}l^\nu,
\;\;\;\;\;\;\;\;\;\;
v\equiv
\eta_{\mu\nu}q^{\mu}_{--}l^\nu,
\end{equation}
and satisfy:
\begin{equation}
u_-=v_+=0
\;\;\;\;\;\;\Longrightarrow\;\;\;\;\;\;u=u(\sigma^+),\;\;\;v=v(\sigma^
-).
\end{equation}
Then, by differentiating equation (2.10) twice, we get:
\begin{equation}
\alpha_{+-}
+u(\sigma_+) v(\sigma_-) e^{-\alpha}=0.
\end{equation}
If the
product $u
(\sigma_+) v(\sigma_-)$ is positive definite, then the following
conformal transformation on the world-sheet metric (2.14):
\begin{eqnarray}
&\alpha(\sigma_+,\sigma_-)=
\hat{\alpha}
(\hat{\sigma}_+,\hat{\sigma}_-)+\frac{1}{2}
\mbox{log}|u
(\sigma_+)||v(\sigma_-)|,&\nonumber\\
&\hat{\sigma}_+=\int\sqrt{|u
(\sigma_+)}|\;d\sigma_+,\;\;\;\;\;\;\;\;
\hat{\sigma}_-=\int\sqrt{|v
(\sigma_-)}|\;d\sigma_-,&
\end{eqnarray}
reduces equation (2.20) to:
\begin{equation}
\alpha_{+-}+
e^{-\alpha}=0,
\end{equation}
which is just the Liouville equation (we skipped the hats).

It must be noticed, however, that for a generic string world-sheet,
the product
$u
(\sigma_+) v(\sigma_-)$ is neither positive nor negative
definite.
In the case that
$u
(\sigma_+) v(\sigma_-)$ is negative, the conformal
transformation
(2.21) reduces equation (2.20) to:
\begin{equation}
\alpha_{+-}-
e^{-\alpha}=0,
\end{equation}
and including also the case when
$u(\sigma_+) v(\sigma_-)=0,$ we conclude that the most
general
equation fulfilled by the fundamental quadratic form
$\alpha$
is:
\begin{equation}
\alpha_{+-}+
Ke^{-\alpha}=0,
\end{equation}
where:
\begin{equation}
K=\left\{ \begin{array}{l}
+1,\;\;\;\;\;\;u(\sigma_+) v(\sigma_-)>0 \\
-1,\;\;\;\;\;\;u(\sigma_+) v(\sigma_-)<0 \\
\;0,\;\;\;\;\;\;\;\;u(\sigma_+)
v(\sigma_-)=0
\end{array}\right.
\end{equation}
Equation (2.24) is either the
Liouville equation ($K=\pm 1$), or the free wave equation $(K=0)$.

Let us
define a potential $V(\alpha)$ by:
\begin{equation}
\alpha_{+-}+\frac{dV
(\alpha)}
{d\alpha}=0,
\end{equation}
so that if $\alpha=\alpha(\tau),$ then
$\;\frac{1}{2}(\dot{\alpha})^2+V
(\alpha)={\mbox{const}}$. Then, it
follows that:
\begin{equation}
V(\alpha)=\left\{ \begin{array}{r}
-e^{-\alpha},\;\;\;\;\;\;K=+1 \\
e^{-\alpha},\;\;\;\;\;\;K=-1\\
\;\;\;\;\;0,\;\;\;\;\;\;\;\;\;\;\;K=0 \end{array}\right.
\end{equation}
The results (2.26)-(2.27) are represented in Fig.1., showing the
different potentials.\\
\\
\\
It is interesting to compare these results
with the analogue results obtained in AdS but
without torsion \cite{san,mik,nes,all}.
In that case,  instead of eq.(2.27), we have found \cite{all}:
\begin{equation}
\tilde{V}(\alpha)=\left\{ \begin{array}{r}
2\sinh\alpha,\;\;\;\;\;\;K=+1 \\
2\cosh\alpha,\;\;\;\;\;\;K=-1\\
\;\;\;\;\;e^\alpha,\;\;\;\;\;\;\;\;\;\;\;K=0 \end{array}\right.
\end{equation}
which is shown in Fig.2. As discussed in more detail in \cite{all},
it
means that for large proper string
sizes (large
$\alpha$), the potential $\tilde{V}(\alpha)$
is always attractive. The positive increasing potential for positive
$\alpha$
in AdS spacetime
prevents the string from growing indefinetely. That is, gravity as
represented by the metric  will (not
surprisingly in AdS) generally tend to contract a large string.

By comparing equations (2.27) and (2.28), we see that the effect of
conformal invariance is to {\it precisely}
cancel the term $e^\alpha$ in the potential. This holds for all three
cases
($K=0,\pm 1$). That is, when including the parallelizing torsion, the
original $\sinh$-Gordon and $\cosh$-Gordon equations reduce to the
Liouville equation (with different signs of the potential), while the
original Liouville equation reduces to the free wave equation.

Thus the physical effect of conformal invariance (represented
via the parallelizing torsion) is to {\it precisely} cancel the
dominant
attractive part of the potential
arising from the metric. In other words, the  parallelizing torsion
generally gives rise to a repulsive
term
$-e^\alpha$ in the potential. The combined effect of gravity and
torsion
eventually gives rise to either
attraction or repulsion, but for large proper string size $\alpha$,
the
potential $V(\alpha)$ vanishes exponentially
in all cases, Fig.1.

On the other hand, for small proper string size
$\alpha$, the potential is not
affected by the parallelizing torsion.

These results complete and generalize results obtained in \cite{vega}
for particular string
configurations (circular strings), and in \cite{all} for the
non-conformally invariant case.\\
\\
\\
Finally, it should be noticed that the general solution of the
Liouville
equation (2.24) is known in closed
form (say, $K=1$):
\begin{equation}
\alpha(\sigma^+,\sigma^-)=\log \left\{ \frac
{(f(\sigma^+)+g(\sigma^-))^2}{2f'(\sigma^+)g'(\sigma^-)} \right\}
\end{equation}
where $f(\sigma^+)$ and $g(\sigma^-)$ are arbitrary functions of the
indicated variables.
The proper string size, $S(\tau)$, is then:
\begin{equation}
S(\tau)=\int d\sigma \; s(\tau,\sigma),
\end{equation}
where, using equations (2.14) and (2.29),
\begin{equation}
s(\tau,\sigma)=\frac{\sqrt{2}}{H}e^{\alpha/2}=
\frac{f(\sigma^+)+g(\sigma^-)}
{H\sqrt{f'(\sigma^+)g'(\sigma^-)}}.
\end{equation}
This is the general solution ($K=1$) to the string size in the
conformally
invariant AdS
background. The full string dynamics in this background is exactly
integrable. However, it
is still a highly non-trivial problem to obtain the explicit
expression for
the coordinates
$q^\mu$, taking into account the constraints (2.12) and the
normalization
condition (2.13).
\section{Examples}
\setcounter{equation}{0}
In this section we consider in detail some illustrative examples of
string
configurations. It
is convenient to first introduce the standard parametrization (see
for
instance \cite{rind})
of
$2+1$ AdS in terms of static coordinates $(t,r,\phi)$:
\begin{eqnarray}
&X=r\cos\phi,\;\;\;\;\;\;U=\frac{1}{H}\sqrt{1+H^2
r^2}\;\cos(Ht),&\nonumber\\
&Y=r\sin\phi,\;\;\;\;\;\;T=\frac{1}{H}\sqrt{1+H^2 r^2}\;\sin(Ht),&
\end{eqnarray}
which automatically fulfils the normalization condition (2.13). Next
we
make the following
ansatz \cite{all}:
\begin{eqnarray}
r&=&r(\xi^1),\nonumber\\
t&=&t(\xi^1)+c_1\xi^2,\\
\phi&=&\phi(\xi^1)+c_2\xi^2,\nonumber
\end{eqnarray}
where $(c_1,c_2)$ are arbitrary constants while $(\xi^1,\xi^2)$ are
the two
world-sheet
coordinates, to be specified later.

The mathematical motivation for this ansatz is that it reduces the
string
equations of motion
(2.9) to ordinary differential equations, as we now show (see also
Ref.\cite{all}). In fact, the equations
(2.9) reduce to:
\begin{eqnarray}
\frac{d^2t}{(d\xi^1)^2}&+&
\frac{2H^2r}{1+H^2r^2}\left(\frac{dt}{d\xi^1}\right)
\left(\frac{dr}{d\xi^1}
\right)
+\frac{2Hr}{1+H^2r^2}\left(\frac{dr}{d\xi^1}\right)c_2=0,\\
\frac{d^2\phi}{(d\xi^1)^2}&+&
\frac{2}{r}\left(\frac{d\phi}{d\xi^1}\right)
\left(\frac{dr}{d\xi^1}\right)
+\frac{2H}{r}\left(\frac{dr}{d\xi^1}\right)c_1=0,\\
\frac{d^2 r}{(d\xi^1)^2}&+&
H^2r(1+H^2r^2)\left(\left(\frac{dt}{d\xi^1}\right)^2-c_1^2\right)-
r(1+H^2r^2)
\left(\left(\frac{d\phi}{d\xi^1}\right)^2-c_2^2\right)\nonumber\\
&-&\frac{H^2r}{1+H^2r^2}\left(\frac{dr}{d\xi^1}\right)^2+
2Hr(1+H^2r^2)
\left(c_2\left(\frac{dt}{d\xi^1}\right)-
c_1\left(\frac{d\phi}{d\xi^1}\right)
\right)=0,\nonumber\\
\end{eqnarray}
while the constraints (2.12) become:
\begin{equation}
(1+H^2r^2)\left(\frac{dt}{d\xi^1}\right)c_1=
r^2\left(\frac{d\phi}{d\xi^1}
\right)c_2,
\end{equation}
\begin{equation}
\frac{1}{1+H^2r^2}\left(\frac{dr}{d\xi^1}\right)^2-
(1+H^2r^2)\left(\left(\frac{dt}{d\xi^1}\right)^2+c_1^2\right)+
r^2\left(\left(\frac{d\phi}{d\xi^1}\right)^2+c_2^2\right)=0.
\end{equation}
The above equations of motion and constraints are consistently
integrated to:
\begin{equation}
\frac{d t}{d\xi^1}=\frac{k_1-Hc_2 r^2}{1+H^2r^2},\;\;\;\;\;\;\;\;
\frac{d \phi}{d\xi^1}=\frac{k_2-Hc_1 r^2}{r^2},
\end{equation}
\begin{equation}
r'^2=\frac{(H^2k_2^2-k_1^2)(c_1^2+2Hc_1k_2)}{r^2k_2^2}
\left(r^2-\frac{k_2^2}{c_1^2+2Hc_1k_2}\right)
\left(r^2+\frac{k_2^2}{H^2k_2^2-k_1^2}\right),
\end{equation}
where the integration constants $(k_1,k_2)$ fulfil:
\begin{equation}
c_1 k_1=c_2 k_2.
\end{equation}
The equations (3.8)-(3.9) can be solved explicitly in closed form in
terms
of trigonometric or
hyperbolic  functions. This is a great simplification compared to the
case
without torsion. In that
case, the solution generally involved elliptic functions \cite{all}.

As for the fundamental quadratic form $\alpha$, we get:
\begin{equation}
e^\alpha=\pm\frac{H^2}{2}\left[ r^2 c_2^2-(1+H^2r^2)c_1^2\right],
\end{equation}
where the sign must be chosen in accordance with eq.(2.14).
Then we get from equations (3.3)-(3.10):
\begin{equation}
\frac{d^2\alpha}{(d\xi^1)^2}\pm
\left[H^2\left( c_2^2-H^2 c_1^2\right)
\left( k_1^2-(c_1+Hk_2)^2\right) \right]
e^{-\alpha}=0.
\end{equation}
This corresponds to equation (2.24) after a constant redefinition of
$\xi^1$. The different values
of $K$ will appear depending on the sign of the square bracket in
(3.12).

In the following subsections, we consider some more explicit examples
to
clarify the physics of the
ansatz (3.2).
\subsection{Circular Strings}
Circular strings are obtained from the above general formalism by
setting
$(\xi^1,\xi^2)=(\tau,\sigma)$, as well as:
\begin{equation}
c_1=0,\;\;\;\;\;\;k_2=0,\;\;\;\;\;\;c_2=1,\;\;\;\;\;\;k_1\equiv E.
\end{equation}
Then equations (3.8)-(3.9) become:
\begin{eqnarray}
\phi&=&\sigma,\nonumber\\
\dot{t}&=&\frac{E-Hr^2}{1+H^2r^2},\\
\dot{r}^2&+&(1+2EH)r^2=E^2. \nonumber
\end{eqnarray}
Here we must take $E\geq 0$ to ensure that $\dot{t}\geq 0$ (the
string is
propagating forward in
time). Then (3.14) describes a circular string oscillating between
$r=0$
and $r=r_{\mbox{max}}$:
\begin{equation}
r_{\mbox{max}}=\frac{E}{\sqrt{1+2EH}}.
\end{equation}
The explicit solution of (3.14), in closed form, is:
\begin{eqnarray}
\phi&=&\sigma,\nonumber\\
Ht&=&\arctan\left(
\frac{1+EH}{\sqrt{1+2EH}}\tan(\sqrt{1+2EH}\;\tau)\right)
-\tau,\nonumber\\
r&=&\frac{E}{\sqrt{1+2EH}}\left| \sin(\sqrt{1+2EH}\;\tau)\right| ,
\end{eqnarray}
where we took initial conditions ($r(0)=0,\;t(0)=0$).  Here,
\begin{equation}
e^\alpha= \frac{H^2r^2}{2}
\end{equation}
and the string size, equations (2.30)-(2.31), is:
\begin{equation}
S(\tau)=2\pi r(\tau).
\end{equation}
These circular
strings have been discussed in
more detail in Ref.\cite{vega}.
\subsection{Stationary Strings}
Stationary strings are obtained from the general formalism by setting
$(\xi^1,\xi^2)=(\sigma,\tau)$, as well as:
\begin{equation}
c_2=0,\;\;\;\;\;\;k_1=0,\;\;\;\;\;\;c_1=1,\;\;\;\;\;\;k_2\equiv L.
\end{equation}
Then equations (3.8)-(3.9) become:
\begin{eqnarray}
t&=&\tau,\nonumber\\
\phi'&=&\frac{L-Hr^2}{r^2},\\
{r'}^2&=&\frac{H^2(1+2HL)}{r^2}\left( r^2+
\frac{1}{H^2}\right) \left(
r^2-\frac{L^2}{1+2HL}\right) .
\nonumber
\end{eqnarray}
It follows that we must have $1+2HL> 0$ to ensure that a region
exists
where ${r'}^2 \geq 0$.
There is also a "turning point" ($r'=0$) at $r=r_{\mbox{min}}$:
\begin{equation}
r_{\mbox{min}}=\frac{|L|}{\sqrt{1+2HL}},
\end{equation}
thus the stationary string stretches out from  $r=r_{\mbox{min}}\;$
to
$r=\infty$.
The explicit solution of (3.18), in closed form, is:
\begin{eqnarray}
t&=&\tau,\nonumber\\
\phi&=&\arctan\left(
\frac{\sqrt{1+2HL}}{{HL}}\tanh(H\sqrt{1+2HL}\;\sigma)\right)
-H\sigma,\nonumber\\
r&=&\sqrt{\frac{(1+HL)^2\;\sinh^2(H\sqrt{1+2HL}\;\sigma)+
H^2L^2}{H^2(1+2HL)}},
\end{eqnarray}
where we took initial conditions ($r(0)=r_{\mbox{min}},\;\phi(0)=0$).
Here,
\begin{equation}
e^\alpha=\frac{1}{2}(1+H^2r^2),
\end{equation}
where a factor $H$ has been absorbed in $\tau$ and $\sigma$.
Then the string size, equations (2.30)-(2.31), is:
\begin{equation}
S(\tau)=S=\int_{-\infty}^{+\infty}\sqrt{1+H^2 r^2}\:d\sigma=\infty.
\end{equation}
Notice that:
\begin{equation}
\phi(-\infty)=\infty,\;\;\;\;\;\;\phi(\infty)=-\infty,
\end{equation}
so that the stationary strings are open infinitely long clockwise
spirals.
An example is shown in Fig.3. Asymptotically
($|\sigma|\rightarrow\infty$), the stationary strings
are standard logaritmic spirals:
\begin{equation}
\left( \begin{array}{cc} X \\
Y \end{array}\right) = \left( \begin{array}{cc} r\cos\phi \\
r\sin\phi \end{array}\right) \sim\; \left(
\begin{array}{cc} \cos(H\sigma) \\
\sin(H\sigma) \end{array}\right)e^{\pm H\sqrt{1+2HL}\;\sigma},
\end{equation}
up to a constant scaling and a rotation. The simplest explicit
example is
obtained for $L=0$:
\begin{equation}
\left( \begin{array}{cc} X \\
Y \end{array}\right) = \left(
\begin{array}{cc} \cos(H\sigma) \\
-\sin(H\sigma) \end{array}\right)\sinh(H\sigma),
\end{equation}
which is shown in Fig.4.

Again it is interesting to compare with the case of stationary
strings in AdS
spacetime, but without torsion \cite{all2}. In that case, the
stationary
strings were of the "hanging string" type, that is, the shape of the
stationary strings was a simple generalization of the shape of a rope
hanging in a constant  Newtonian potential. Here, in the presence of
torsion,
the stationary strings are instead of the "spiralling string" type.
It follows that the effect of torsion is somewhat similar to the
effect of
rotation in the metric: The effect of rotation in the metric on the
shape of
stationary strings was first investigated in Ref.\cite{zel}. It was
shown
that stationary strings in the Schwarzschild background are of the
"hanging string" type, while in the Kerr background, stationary
strings
could also
be of the "spiralling string" type. Thus, we have seen that
the effect of torsion in AdS spacetime is quite
similar.

Another effect of the torsion on stationary strings in AdS spacetime
concerns the multi-string property: In AdS spacetime without torsion
\cite{all2}, it was shown that the solution corresponding to the
ansatz (3.2), (3.17) actually  describes a multi-string, that is, one
single world-sheet, determined by one set of initial conditions,
describes
a finite or even an infinite number of different and independent
stationary
strings. Here, in the presence of torsion, the multi-string property
is lost
for stationary strings: for $\sigma\in\;]-\infty,\;\infty[\;\;$ the
solution (3.20) describes only one stationary string.
\section{Strings in the BH-AdS background with Torsion}
\setcounter{equation}{0}
In Section 3, we have been concerned with {\it global} 2+1 AdS
spacetime.
However, the general results obtained in Section 2
hold for any parametrization of the SL(2,R) WZWN model. That is to
say,
everything in Section 2 is valid also for strings in the 2+1 black
hole
anti de Sitter spacetime (BH-AdS) \cite{ban}. The 2+1 BH-AdS
spacetime
is obtained by
replacing the SL(2,R)
parametrization (3.1) by \cite{ban2}:
\begin{eqnarray}
a&=&\sqrt{\frac{r^2-r_-^2}{r_+^2-r_-^2}}\;
e^{H(r_+\phi-Hr_- t)}\:,\nonumber\\
b&=&\sqrt{\frac{r^2-r_-^2}{r_+^2-r_-^2}}\;
e^{-H(r_+\phi-Hr_- t)}\:,\nonumber\\
c&=&\sqrt{\frac{r^2-r_+^2}{r_+^2-r_-^2}}\;
e^{H(Hr_+ t-r_- \phi)}\:,\nonumber\\
d&=&-\sqrt{\frac{r^2-r_+^2}{r_+^2-r_-^2}}\;
e^{-H(Hr_+\phi-r_- \phi)}\:,
\end{eqnarray}
where we used the notation of eq.(2.2). In these expressions $r_\pm$
are the
outer
and inner horizons:
\begin{equation}
r_\pm^2=\frac{M}{2H^2}(1\pm\sqrt{1-H^2J^2/M^2}\;),
\end{equation}
where $(M,J)$ represent the mass and angular momentum of the black
hole,
respectively. Finally we used the notation $H^{-1}=l$ (the length
scale) for
comparison with sections 2, 3.
Notice also that eq.(4.1) is only valid for $r>r_+$, but analogous
expressions
hold in the other regions. For more details about the BH-AdS
spacetime, we
refer the readers to the original papers \cite{ban,ban2}.

It is now straightforward to perform the analysis of circular and
stationary
strings in the background of 2+1 BH-AdS with torsion, c.f. the
analysis of
Section 3, so here we just give the main results.
\subsection{Circular Strings}
The equations of motion and constraints for circular strings are
solved by:
\begin{equation}
t=\int^\tau\frac{E-Hr^2}{H^2r^2-M+J^2/4r^2}\;d\tau\:,
\end{equation}
\begin{equation}
\phi=\sigma+\int^\tau
\frac{J(E-Hr^2)}{2r^2(H^2r^2-M+J^2/4r^2)}\;d\tau\:,
\end{equation}
\begin{equation}
\dot{r}^2+V(r)=0\;\;\;\;\Leftrightarrow\;\;\;\;\tau=\pm \int^r
\frac{dr}{\sqrt{-V(r)}}\:,
\end{equation}
where:
\begin{equation}
V(r)=-(M-2EH)r^2-(E^2-J^2/4),
\end{equation}
and $E$ is an integration constant. Here:
\begin{equation}
e^\alpha=\frac{H^2r^2}{2},
\end{equation}
and we have:
\begin{equation}
\ddot{\alpha}+H^2(E^2-J^2/4)e^{-\alpha}=0.
\end{equation}

Now taking into account that $M>0$, $H>0$,
$|J|\leq M/H$ as well as the {\it physical} requirement that the
string
propagates forward in time (at least outside the horizon), we must
have:
\begin{equation}2EH-M>0,
\end{equation} which also implies that:
\begin{equation}
E^2>J^2/4.
\end{equation}
It follows that the potential is a monotonically increasing function,
and that the circular string contracts from $r=r_{\mbox{max}}$ to
$r=0$, where:
\begin{equation}
r_{\mbox{max}}=\sqrt{\frac{E^2-J^2/4}{2EH-M}}.
\end{equation}
Depending on $E$, the maximal string radius can be larger or smaller
than
$r_+$. In the first case, the string will initially be outside the
horizon, but
will then contract, fall into it and collapse into $r=0$. In the
latter case,
the string is always
inside the horizon and collapses into $r=0$.

In the case without torsion \cite{all3}, the potential $V(r)$ was
quartic in
$r$ and the solutions involved elliptic functions. As a consequence,
the
possibility $E^2<J^2/4$ (Sinh-Gordon sector) also appeared, and a
potential
barrier between the inner horizon $r_-$ and $r=0$, preventing the
string from
collapsing into $r=0$, was present.
\subsection{Stationary Strings}
The equations of motion and constraints for stationary strings are
solved by:
\begin{equation}
t=\tau-\int^\sigma
\frac{J(L-Hr^2)}{2r^2(H^2r^2-M+J^2/4r^2)}\;d\sigma\:,
\end{equation}
\begin{equation}
\phi=\int^\sigma
\frac{(L-Hr^2)(H^2r^2-M)}{r^2(H^2r^2-M+J^2/4r^2)}\;d\sigma\:,
\end{equation}
\begin{equation}
{r'}^2+U(r)=0\;\;\;\;\Leftrightarrow\;\;\;\;\sigma=\pm \int^r
\frac{dr}{\sqrt{-U(r)}}\:,
\end{equation}
where:
\begin{equation}
U(r)=(H^2r^2-M)\left[ \frac{L^2-J^2/4}{r^2}+(M-2LH)\right],
\end{equation}
and $L$ is an integration constant. Here:
\begin{equation}
e^{\alpha}=\frac{1}{2}(H^2r^2-M),
\end{equation}
and we have:
\begin{equation}
\alpha''+H^2\left[ M(2HL-M)-H^2(L^2-J^2/4)\right] e^{-\alpha}=0.
\end{equation}

Now taking into account that $M>0$, $H>0$,
$|J|\leq M/H$ as well as the {\it physical} requirement that the
stationary
string  (at least a part of) must be outside the static limit, we
must have:
\begin{equation}
2LH-M>0,
\end{equation}
which also implies that:
\begin{equation}
L^2>J^2/4.
\end{equation}
It follows that the stationary string stretches out to infinity, but
there is a
"turning point"($r'=0$) at $r=r_{\mbox{min}}$:
\begin{equation}
r_{\mbox{min}}=\sqrt{\frac{L^2-J^2/4}{2LH-M}}.
\end{equation}
Depending on $L$, the turning point can be outside or inside the
static limit
$r_{st}=\sqrt{M}/H$. In the first case, the string will be of
"hanging string"
type, with both ends at infinity, while in the latter case it will be
of
"spiralling string" type with one end at infinity, crossing the
static limit
and spiralling into the black hole. In the limiting case, when
$r_{\mbox{min}}$
is equal to $r_{st}$, corresponding to:
\begin{equation}
L=\frac{M}{H}\pm\frac{|J|}{2},
\end{equation}
the solution just fulfills the free wave equation, interpolating
between the
"hanging string" and the "spiralling string" types.

\section{Conclusion}
Using a physical approach, working directly with the classical string
equations
of motion and the proper string size, we reduced the SL(2,R)
WZWN model to Liouville
theory. This allowed us to extract the precise physical
effects of the parallelizing torsion on the generic string dynamics.
We
showed that the parallelizing torsion, corresponding to conformal
invariance, generally led to repulsion. In fact, the parallelizing
torsion
gives rise to a repulsive term that {\it precisely} cancels the
dominant
attractive term arising from the metric. As a consequence, the
sinh-Gordon
and cosh-Gordon
sectors of the non-conformally invariant AdS background reduce to the
Liouville equation
(with different signs of the potential), while the original Liouville
sector reduces to the
free wave equation. Thus, the dynamics of the classical large size
strings
is affected by
the torsion, but most of the string size behaviour (intermediate and
small
sizes) is quite
the same. We also gave the general solution to the proper string
size.

We then analysed in detail the circular and stationary strings in the
AdS
spacetime and in the 2+1 BH-AdS spacetime, both  with parallelizing
torsion.
These  results confirmed our generic
results (as they should), and we compared with the case of vanishing
torsion. In particular, it was shown that the effect of torsion on
the
stationary strings is quite similar to the effect of rotation in
the metric.

\setcounter{equation}{0}

\newpage
\begin{figure}
\centerline{\epsfig{figure=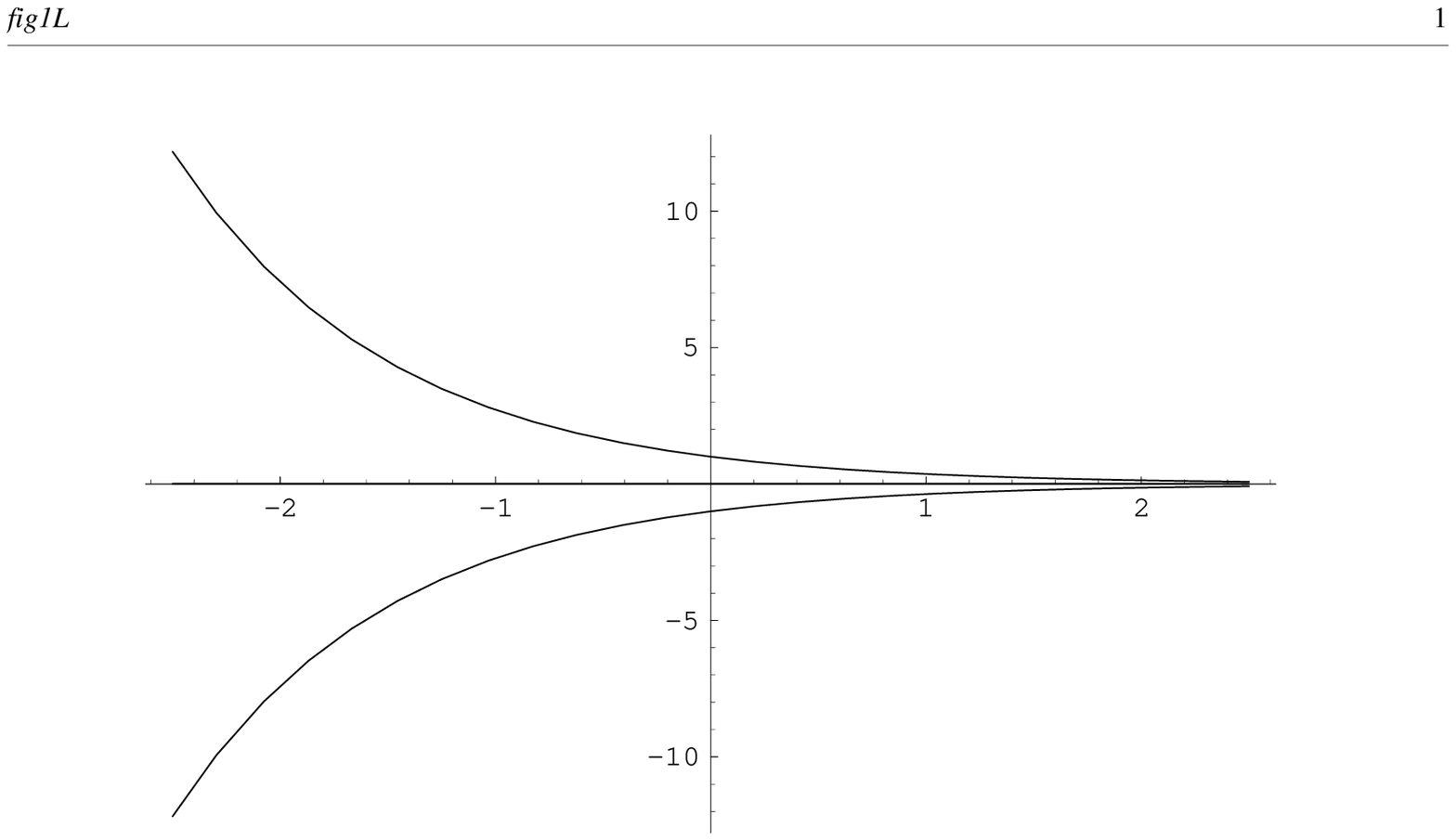, height=10cm,angle=0}}
\caption {The potential (2.27). The three curves represent, from above,
K=-1, K=0 and K=1,
respectively.}
\end{figure}
\begin{figure}
\centerline{\epsfig{figure=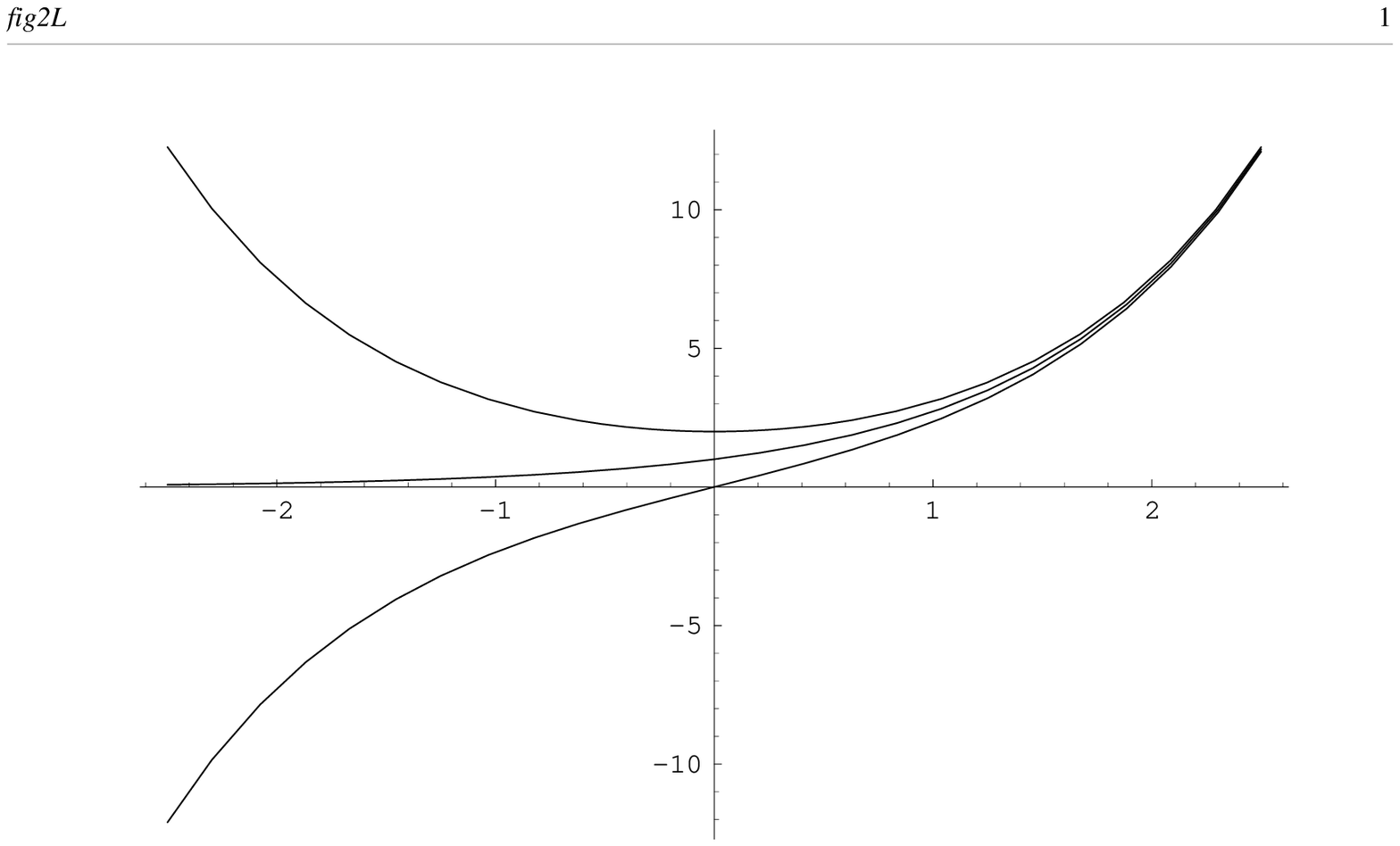, height=10cm,angle=0}}
\caption {The potential (2.28). The three curves represent, from above,
K=-1, K=0 and K=1,
respectively.}
\end{figure}
\begin{figure}
\centerline{\epsfig{figure=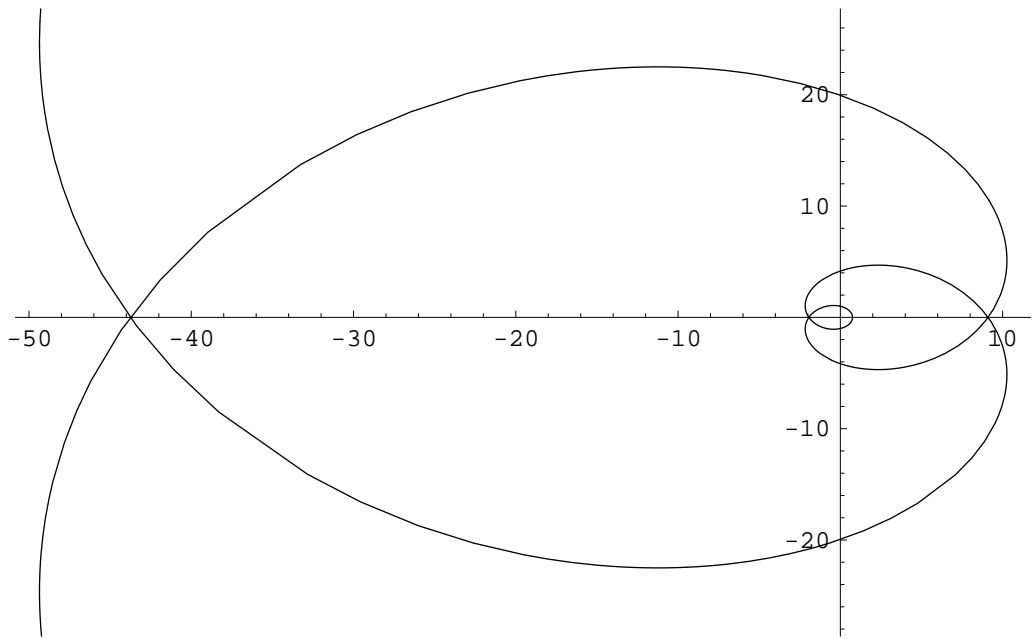, height=10cm,angle=0}}
\caption {The stationary string (3.20) corresponding to the case HL=-3/8.}
\end{figure}
\begin{figure}
\centerline{\epsfig{figure=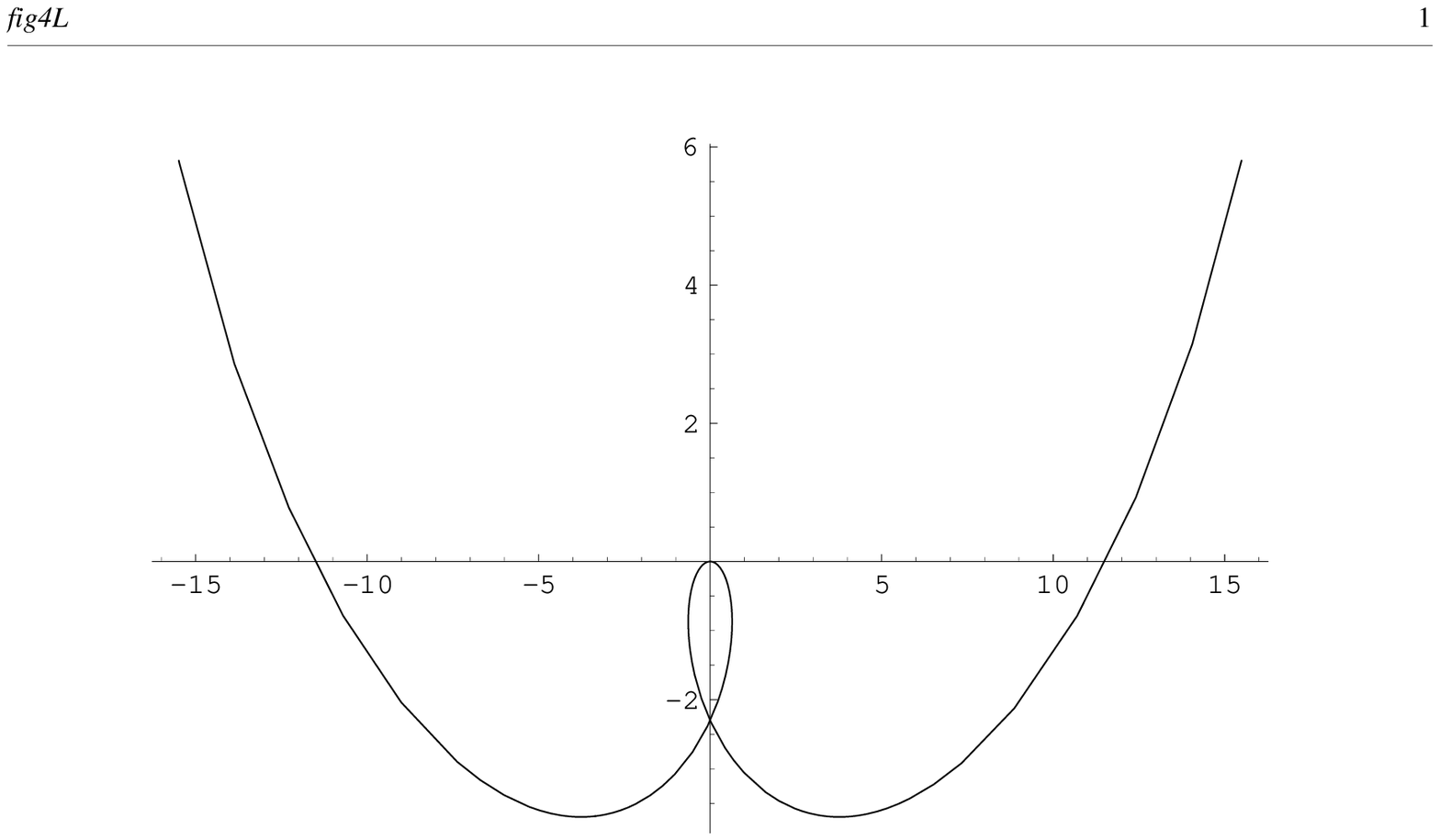, height=10cm,angle=0}}
\caption {The stationary string (3.27) corresponding to the case HL=0.}
\end{figure}

\newpage
\begin{thebibliography}{11}
\bibitem{bal}J. Balog, L. O'Raifeartaigh, P. Forgacs and A. Wipf,
Nucl.
Phys. {\bf B325} (1989) 225.
\bibitem{pet}P. Petropoulos, Phys. Lett. {\bf B236} (1990) 151.
\bibitem{nem}I. Bars and D. Nemeschansky, Nucl. Phys. {\bf B348}
(1991) 89.
\bibitem{hwa}S. Hwang, Nucl. Phys. {\bf B354} (1991) 100.
\bibitem{rai}L. O'Raifeartaigh, V.V. Sreedhar, "Conformally Invariant
Path
Integral Formulation of
the  Wess-Zumino-Witten $\rightarrow$ Liouville Reduction". Preprint,
hep-th/9709143.
\bibitem{san}H.J. de Vega and N. S\'{a}nchez, Phys. Rev. {\bf D47}
(1993) 3394.
\bibitem{all}A.L. Larsen and N. S\'{a}nchez, Phys. Rev. {\bf D54}
(1996) 2801.
\bibitem{mik}H.J. de Vega, A.V. Mikhailov and N. S\'{a}nchez, Mod.
Phys. Lett. {\bf A29} (1994) 2745; Teor. Mat. Fiz. {\bf 94} (1993)
232.
\bibitem{nes}B.M. Barbashov and V.V. Nesterenko, Commun. Math. Phys.
{\bf 78} (1981) 499.
\bibitem{vega} H.J. de Vega, A.L. Larsen and  N. S\'{a}nchez,
 Phys. Rev.
{\bf D58} (1998) 026001 .
\bibitem{zel}V.P. Frolov, V. Skarzhinski, A. Zelnikov, O. Heinrich,
Phys. Lett.
{\bf B224} (1989) 255.
\bibitem{wit}E. Witten, Commun. Math. Phys. {\bf 92} (1984) 455.
\bibitem{hor}J. Horne and G.T. Horowitz, Nucl. Phys. {\bf B368}
(1992) 444.
\bibitem{rind}W. Rindler, "Essential Relativity" (Springer Verlag,
New
York, 1979). Chapter 8.11.
\bibitem{all2}A.L. Larsen and N. S\'{a}nchez, Phys. Rev. {\bf D51}
(1995) 6929.
\bibitem{ban}M. Banados, M. Henneaux, C. Teitelboim and J. Zanelli,
Phys. Rev.
{\bf D48}.
(1993) 1506.
\bibitem{ban2}M. Banados, C. Teitelboim and J. Zanelli, Phys. Rev.
Lett.
{\bf 69} (1992) 1849.
\bibitem{all3}A.L. Larsen and N. S\'{a}nchez, Phys. Rev. {\bf D50}
(1994) 7493.
\end{thebibliography}
\end{document}